\documentclass[aps,prl,twocolumn,showpacs,superscriptaddress]{revtex4-1}

\usepackage{graphicx}
\usepackage{amssymb}

\begin{document}
\title{New Superconductivity Dome in LaFeAsO$_{1-x}$F$_{x}$ Accompanied by Structural Transition}


\author{J. Yang}
\affiliation
{Institute of Physics and Beijing National Laboratory for Condensed Matter Physics,
\\Chinese Academy of Sciences, Beijing 100190, P. R. China}

\author{R. Zhou}
\affiliation
{Institute of Physics and Beijing National Laboratory for Condensed Matter Physics,
\\Chinese Academy of Sciences, Beijing 100190, P. R. China}

\author{L. L. Wei}
\affiliation
{Institute of Physics and Beijing National Laboratory for Condensed Matter Physics,
\\Chinese Academy of Sciences, Beijing 100190, P. R. China}

\author{H. X. Yang}
\affiliation
{Institute of Physics and Beijing National Laboratory for Condensed Matter Physics,
\\Chinese Academy of Sciences, Beijing 100190, P. R. China}

\author{J. Q. Li}
\affiliation
{Institute of Physics and Beijing National Laboratory for Condensed Matter Physics,
\\Chinese Academy of Sciences, Beijing 100190, P. R. China}

\author{Z. X. Zhao}
\affiliation
{Institute of Physics and Beijing National Laboratory for Condensed Matter Physics,
\\Chinese Academy of Sciences, Beijing 100190, P. R. China}

\author{Guo-qing Zheng}
\email
{gqzheng123@gmail.com}
\affiliation
{Institute of Physics and Beijing National Laboratory for Condensed Matter Physics,
\\Chinese Academy of Sciences, Beijing 100190, P. R. China}
\affiliation
{Department of Physics, Okayama University, Okayama 700-8530, Japan}



\date{\today}







\begin{abstract}
High temperature superconductivity is often found in the vicinity of antiferromagnetism.
This is also true in LaFeAsO$_{1-x}$F$_{x}$  ($x \leq$ 0.2) and many other iron-based superconductors,  which leads to proposals that superconductivity is mediated by  fluctuations associated with the nearby magnetism. Here we report the discovery of a new superconductivity dome without low-energy magnetic fluctuations in LaFeAsO$_{1-x}$F$_{x}$ with 0.25$\leq x \leq$0.75, where the maximal critical temperature $T_c$ at $x_{opt}$ = 0.5$\sim$0.55 is even higher than that at $x \leq$ 0.2. By nuclear magnetic resonance and Transmission Electron Microscopy, we show that a C4 rotation symmetry-breaking structural transition takes place for $x>$ 0.5 above $T_c$. Our results point to a new paradigm of high temperature superconductivity.


\end{abstract}

\maketitle


In  conventional superconductors, 
the glue binding two electrons to form an electron pair  (Cooper pair) is phonon, the quantization of  lattice vibrations.
In  copper-oxide high temperature superconductors or heavy-fermion compounds, superconductivity emerges in the vicinity of antiferromagnetism,  therefore it is  believed by many that magnetic interactions (fluctuations) associated with the close-by  magnetically-ordered phase mediate the electron pairing \cite{cuprates,Lonzarich}. 
Iron pnictides  are a new class of high temperature superconductors \cite{Hosono,Ren}.
Carrier-doping by element substitution \cite{Hosono,Johrendt,Safet,ZhouRui} or oxygen defect \cite{Ren-O} 
suppress  the magnetic order  and superconductivity appears. 
In the carrier-concentration range where superconductivity emerges, strong low-energy magnetic (spin) fluctuations survive \cite{Ning,Oka,Li}. This naturally leads to many proposals that  superconductivity in the iron pnictides is also due to spin fluctuations associated with the nearby magnetic phase \cite{Mazin,Kuroki,Scalapino}.

The prototype iron-pnictide LaFeAsO becomes superconducting when the magnetic order is
suppressed upon replacing a part of oxygen (O)  by fluorine (F). The superconducting transition temperature $T_c$ forms a  dome shape as a function of F-content, with the highest $T_c$ = 27 K at $x$ = 0.06 where strong low-energy spin fluctuations were found by nuclear magnetic resonance (NMR) measurements \cite{Oka};  
the F-content could not exceed 0.2 \cite{Hosono,Luetkens,Oka}.
The high-pressure synthesis technique has proved to be powerful in making high F-concentration compounds \cite{Ren}, and the F-content was able to reach $x$ = 0.6 by this technique~\cite{Lu}. Recently, it was reported that, with high-pressure synthesis technique, hydrogen can also be doped to a high rate of $x$ = 0.53~\cite{Iimura}.
In this work, we succeeded in synthesizing a series of high-doping samples of LaFeAsO$_{1-x}$F$_{x}$ with 0.25 $\leq x \leq$ 0.75 by high-pressure synthesis technique. The $T_c$ forms a new dome shape peaked at $x_{opt}$ = 0.5$\sim$0.55 with a maximal $T_c$ = 30 K below which diamagnetism appears (the electrical resistivity starts to drop at $T_c^{\rho}\sim$40 K). 
The physical properties are completely different from the first dome.
Over the entire new dome, there is  no indication of low-energy magnetic fluctuations as evidenced by 
the spin-lattice relaxation rate ($1/T_1$) divided by temperature ($T$), $1/T_1T$. Instead,   above the dome and for $x>$ 0.5,  we find a new type of phase transition  below which  a four-fold  symmetry (C4 rotation symmetry) is broken.

The polycrystalline samples of LaFeAsO$_{1-x}$F$_{x}$ (with nominal fluorine content $x$ = 0.25, 0.3, 0.4, 0.5, 0.55, 0.6, 0.65, 0.7 and 0.75) were prepared by the high pressure synthesis method. In the first step, the precursor LaAs powder was obtained by reacting La pieces (99.5\%) and As powders (99.999\%). 
In the second step, the starting materials LaAs, Fe (99.9\%), Fe$_{2}$O$_{3}$ (99.9\%) and FeF$_{2}$ (99.9\%) were mixed together according to the nominal ratio 
and sintered in a six-anvil high-pressure synthesis apparatus under a pressure of 6 GPa at 1250$^{\circ}$C for 2-4 hours. After sintering, the sample was quickly quenched to room temperature and then the pressure was released. This rapid quenching process was not used in the previous works \cite{Ren,Lu}, but is important 
 to keep the meta-stable phase which cannot be  formed at ambient pressure.


\begin{figure}[htbp]
\includegraphics[width=9cm]{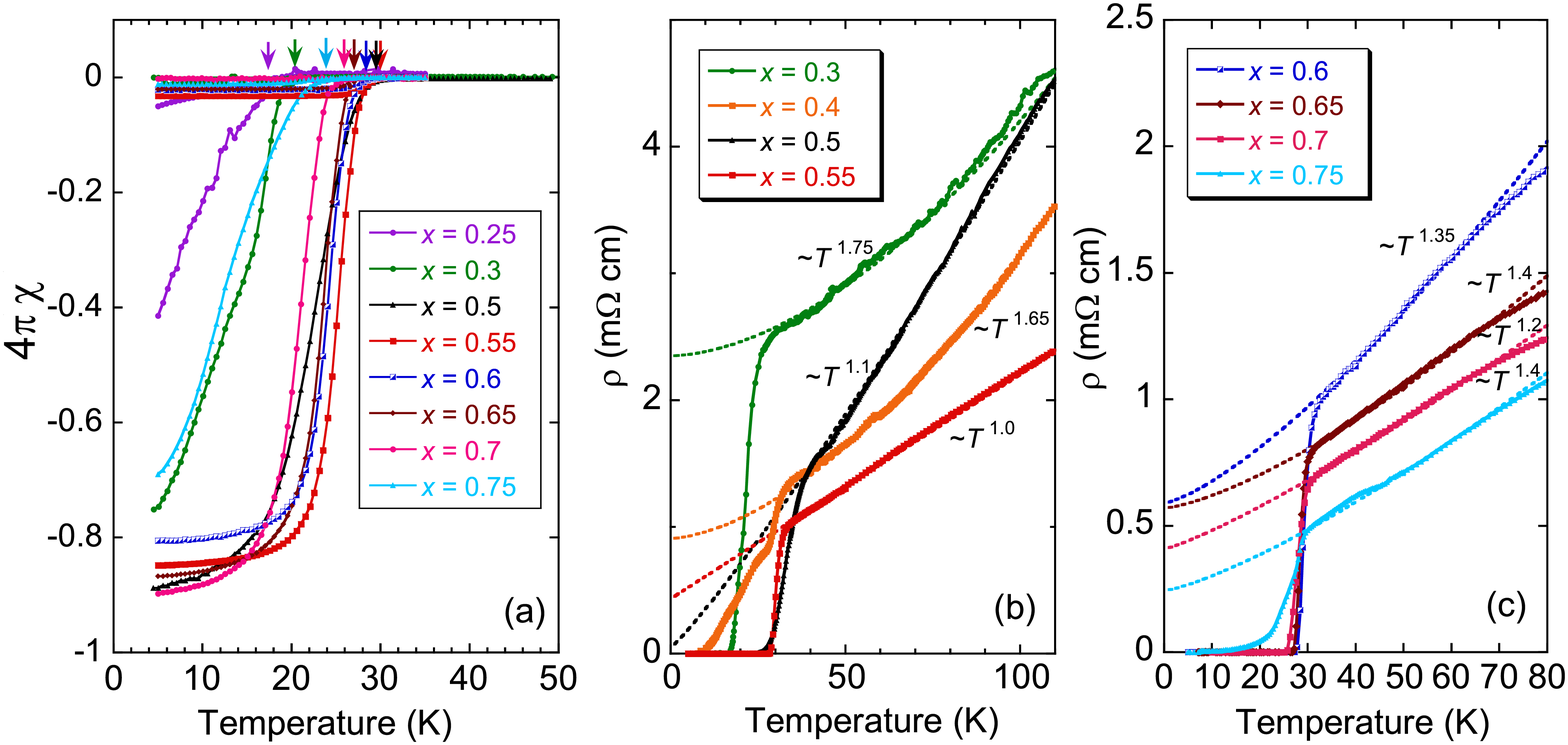}
\caption{The electrical resistivity and DC susceptibility of LaFeAsO$_{1-x}$F$_{x}$. (a) The  DC susceptibility data measured with an applied magnetic field of 10 Oe. The arrows indicate the on-set temperature of the diamagnetism for each sample, which was defined as $T_c$.
(b, c) The low temperature electrical resistivity. 
 The dashed lines are the fittings to $\rho(T) = \rho_0 + A T^{n}$ 
over the temperature range shown. }
\label{RT_chi}
\end{figure}

Powder X-ray diffraction (XRD) with Cu K$\alpha$ radiation were performed at room temperature to characterize the phase and structural parameters.
The temperature dependence of resistivity were measured by a standard four-probe method. 
The $T_c$ was determined by both
DC susceptibility using a superconducting quantum interference device (Quantum Design) and AC susceptibility using an $in~situ$ NMR coil.
For NMR measurements, the samples were crushed into fine powders with each grain's diameter less than 50$\mu$m. The samples were then put in the magnetic field of 12 T at room temperature. Mechanical vibration was added to the sample so that the $a$-axis of the grains are aligned parallel to the magnetic field direction. The nucleus $^{75}$As has a nuclear spin $I$ = 3/2 and the nuclear gyromagnetic ratio $\gamma$ = 7.2919 MHz/T. The NMR spectra were obtained by scanning the RF frequency and integrating the spin echo at a fixed magnetic field $H_{0}$. The spin-lattice relaxation time $T_{1}$ was determined by using the saturation-recovery method.  The nuclear magnetization $M$ can be fitted to $1-M(t)/M(\infty) = 0.1exp(-t/T_{1}) + 0.9exp(-6t/T_{1})$ expected for the central transition peak, where $M(t)$ is the nuclear magnetization at time $t$ after
the single saturation pulse \cite{Narath}. For TEM observations, a JEOL 2100F TEM  equipped with cooling (below $T$=300 K) or heating sample holders (above $T$=300 K), was used for investigating the structural properties of these materials.



\begin{figure}
\includegraphics[width=7cm]{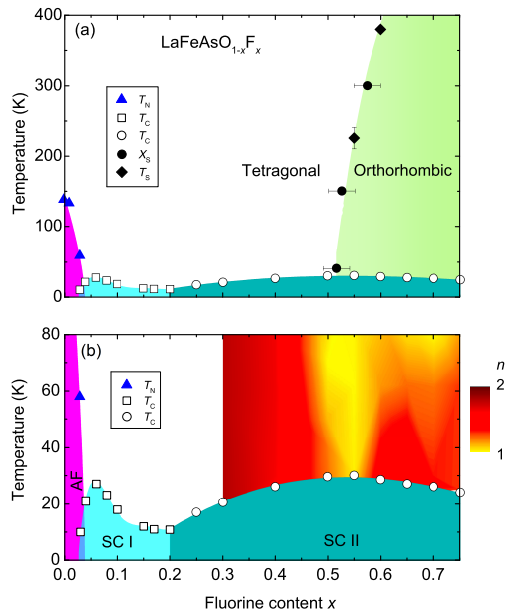}
\caption{ The obtained phase diagram of LaFeAsO$_{1-x}$F$_{x}$. AF denotes the antiferromagnetically ordered phase, SCI and SCII denote the superconducting phases obtained by conventional solid-state and high-pressure  synthesis methods, respectively. (a) Doping dependence of the magnetic ($T_N$), superconducting ($T_c$) and structural transition temperatures ($T_s$).  
The $T_N$ and $T_c$ in SC I are taken from ref. \cite{Oka}. 
The $T_s$ for $x$ = 0.55 and 0.6 was obtained from NMR and TEM measurements, respectively. $x_s$ is the critical F-concentration at which the 
structure transition takes place.  
(b) The evolution of the exponent $n$ obtained by fitting the electrical resistivity to a  relation $\rho (T) = \rho_0 + A T^{n}$.
}
\label{phase_diagram}
\end{figure}

Figure \ref{RT_chi}(a) shows the DC susceptibility data.  The $T_c$ is defined as the onset of the diamagnetism in this paper, which reaches the highest value of 30 K at $x$ = 0.55. The superconducting volume fractions of these samples, as shown in Fig. \ref{RT_chi}(a), indicate the bulk nature of the superconductivity, which was further
assured by a clear decrease of $^{75}$As spin-lattice relaxation rate  below $T_c$ (see below).
Figure \ref{RT_chi}(b)(c) show the temperature dependence of the electrical resistivity $\rho$ for LaFeAsO$_{1-x}$F$_{x}$ (0.3 $\leq x \leq$ 0.75). In the literatures, superconducting critical temperature  was often defined as the temperature below which $\rho$ starts to drop. In such a definition, the highest critical temperature is realized at  $x$=0.5 with $T_c^{\rho}\sim$40 K.
 The resistivity data can be fitted by the equation $\rho = {{\rho }_{0}}+A{{T}^{n}}$ (data over the whole temperature range, see ref.~\cite{Supple}).
For a conventional metal described by Landau Fermi liquid theory, the exponent $n$ = 2 is expected. However, we find $n$ $<$ 2 for all F-concentrations, which is 
suggestive of non-Fermi liquid behavior. Most remarkably, a $T$-linear behavior ($n$ = 1) is  observed for $x$=0.5 over the range of  60 K$< T <$ 110 K,
and for $x$ = 0.55 over the range of 33 K $< T <$ 140 K.
The evolution with Fluorine content $x$ of the exponent $n$, which was obtained by a sliding power law, and $T_c$ are shown in Fig. \ref{phase_diagram}. Note that  neither F-doping nor oxygen-deficiency was able to go beyond $x=$ 0.2 in the past, and the superconductivity in  the high-doping region  is unprecedented.

The crystal symmetry transition  is probed by measuring a change in the electric field gradient (EFG) tensors, $V_{\alpha\alpha}=\frac{\partial^{2}V}{\partial \alpha^{2}}  $ ($\alpha=x,y, z$), at the $^{75}$As and $^{139}$La sites, where $V$ is the electrical potential. The EFG is related to the observables called nuclear quadrupole resonance (NQR) frequency tensors $\nu_{\alpha}=\frac{eQ}{4I(2I-1)}V_{\alpha\alpha}$, where $I$ is the nuclear spin and $Q$ is the nuclear quadrupole moment.
Let us use the following  EFG asymmetry parameter   to describe the  in-plane anisotropy.
 \begin{equation}
 \eta  = \frac{\nu_{x}-\nu_{y}}{\nu_{x}+\nu_{y}}
\end{equation}
When the four-fold symmetry exists in the $ab$-plane, $V_{xx}$=$V_{yy}$ so that  $\eta$ = 0. When the four-fold symmetry is broken, then $V_{xx}$ and $V_{yy}$ are not identical so that  $\eta$ becomes finite. By $^{75}$As NMR and $^{139}$La nuclear quadrupole resonance (NQR),
We find that $\eta$ changes from 0 to a finite value below the structural transition temperature $T_s$ = 40$\sim$380 K for compounds with $x>$0.5.
Figure \ref{phase_diagram}(a) shows the phase diagram that depicts the new superconductivity dome and the symmetry-breaking structural phase transition boundary. 

\begin{figure}
\includegraphics[width=9cm]{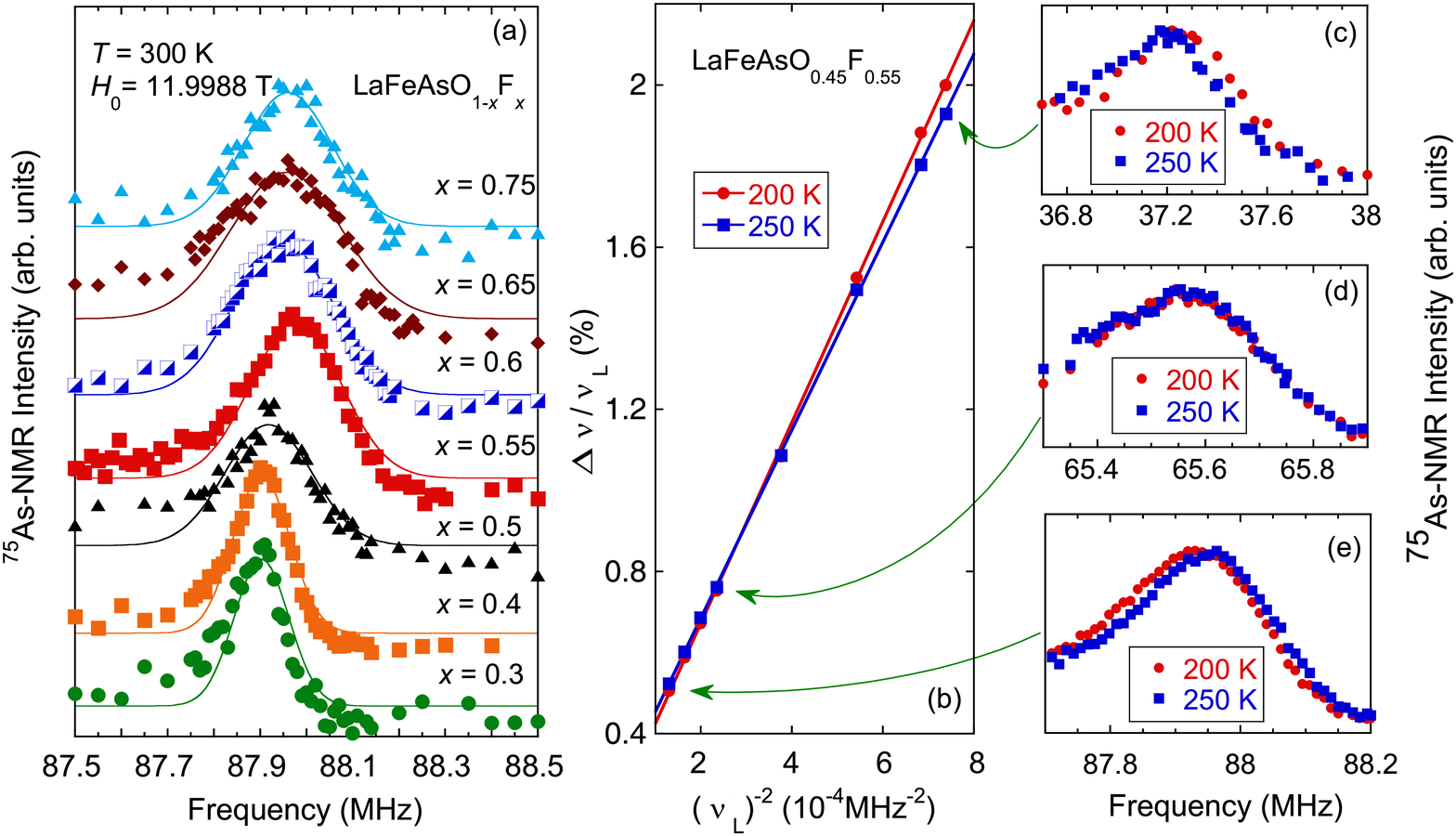}
\caption{ $^{75}$As NMR spectroscopy and the procedure to obtain the quantity $\nu _c\sqrt{1 +  \frac{\eta ^2}{9}}$.
(a) $^{75}$As NMR central transition peak  with $H_{0}$ $\|$ $ab$-plane  for magnetically oriented powder.
The peak corresponds to the central transition (the $m$ = -1/2 $\leftrightarrow$ 1/2 transition).
(b) An example of the plot of  $\Delta\nu$/$\nu_{L}$ against $\nu _L^{ - 2}$ for different $H_{0}$.  The quantities  $K$ and $\nu _c^2(1   + \frac{\eta ^2}{9})$ are obtained  through the intersection with the vertical axis and the slope of the $\Delta\nu$/$\nu_{L}$ versus $\nu _L^{ - 2}$ line, respectively. 
(c-e) Demonstration of  the NMR spectra  at $T$ = 200 K and 250 K for three representative fields $H_{0}$ = 4.9983 T, 8.9232 T and 11.9988 T. 
}
\label{KvsrHandNMRspectra}
\end{figure}

Below we elaborate how the C4 symmetry-breaking structural phase transition is identified. Taking advantage of the fact that the magnetic susceptibility  in the $ab$-plane is larger than that along the $c$-axis\cite{Matano},
 we obtain the $^{75}$As NMR spectra  corresponding to $H_{0}$ $\|$ $a (b)$-axis. 
Figure \ref{KvsrHandNMRspectra}(a) shows the  $^{75}$As NMR  spectra of the magnetically-aligned grains 
under an applied magnetic field $H_{0}$ = 11.9988 T.
The total nuclear spin Hamiltonian is written as $\mathcal{H}$ = $\mathcal{H}_{Z}$ + $\mathcal{H}_{Q}$, where $\mathcal{H}_{Z}=H_0(1+K)$ is the Zeeman interaction with $K$ being the Knight shift due to hyperfine interaction, and
$\mathcal{H}_{Q}=\frac{eV_{zz}Q}{4I(2I-1)} ( (3I_z^2-I^2)+\frac{1}{2}\eta(I_{+}^2+I_{-}^2) )$ is due to the nuclear quadrupole interaction.
The sizable $\mathcal{H}_{Q}$ for  $^{75}$As ($I$=3/2) has a significant perturbation effect on the $m$ = 1/2 $\leftrightarrow$ -1/2 transition
(central transition), leading to a shift of this transition line. When a magnetic field is applied along the principal axis, say, $z$, the resonance frequency shift   is given as  \cite{Abragam},
\begin{equation}
 \Delta \nu  = \nu_{res} - \nu _{L} =  K{\gamma _{N}}{H_{0}} + \frac{{{{({\nu_{x}} - {\nu_{y}})}^2}}}{{12(1+K){\nu _{L}}}}
\end{equation}
Here  $\nu _{L}=\gamma_N H_0$ is the Larmor frequency and $\nu_{res}$ is the observed resonance frequency.
In undoped LaFeAsO at high temperature, the principal axes $x, y, z$ coincide with the crystal $a$-, $b$-, $c$-axis \cite{Imai-LaFeAsO}~.

When a four-fold symmetry is broken so that  $\eta$ is nonzero,
 the second term of the above equation becomes
$\frac{{3\nu _c^2}}{{16(1+K){\nu _{L}}}}(1 - \frac{2}{3}\eta  + \frac{{{\eta ^2}}}{9})$ for  $H_{0}$ $\|$ $a$-axis ,
 and   it becomes
$\frac{{3\nu _c^2}}{{16(1+K){\nu _{L}}}}(1 + \frac{2}{3}\eta  + \frac{{{\eta ^2}}}{9})$ for $H_{0}$ $\|$ $b$-axis.
 Therefore, for a twined single crystal or a magnetically-aligned powder sample, the averaged resonance frequency shift can be written as
\begin{equation}
\frac{{3\nu _c^2}}{{16(1+K){\nu _{L}}}}(1 +  \frac{{{\eta ^2}}}{9})
\end{equation}
%

The quantity $\nu _c\sqrt{1 +  \frac{\eta ^2}{9}}$ was obtained 
from the slop of the plot $\Delta\nu$/$\nu_{L}$ against $\nu _L^{ - 2}$ according to eq. (2) and (3). Figure \ref{KvsrHandNMRspectra}(b-e) show an example of such procedure for the compound of $x$ = 0.55.

\begin{figure*}
\includegraphics[width=12cm]{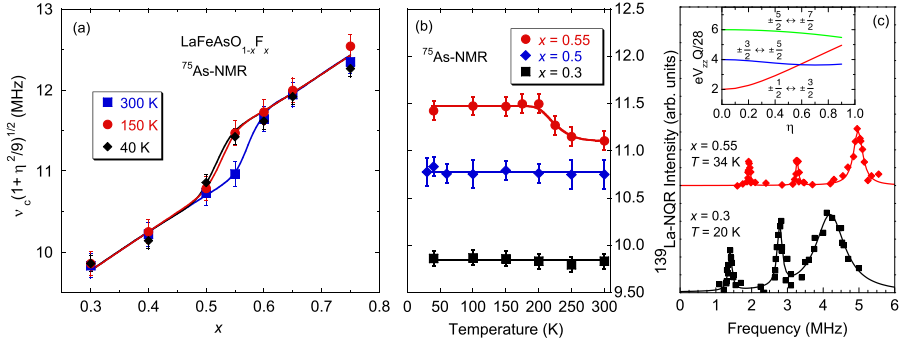}
\caption{ Evidence for structural phase transition from electric field gradient tensors obtained by NMR and NQR.
 (a) The quantity ${\nu _c}\sqrt {1 + {\eta ^2}/9}$ plotted against fluorine content $x$  at $T$=300, 150 and 40 K.
(b) ${\nu _c}\sqrt {1 + {\eta ^2}/9}$ as a function of  temperature for $x$ = 0.3, 0.5 and  0.55.
The solid curves indicate the variation trend of the quantities.
(c) The three $^{139}$La-NQR lines for $x$ = 0.3 are equally spaced, indicating $\eta$ = 0, while those   for  $x$ = 0.55 are not, indicating a finite $\eta$. 
The inset shows the evolution of the three transition frequencies with increasing $\eta$.}
\label{nu_all}
\end{figure*}

Figure \ref{nu_all}(a) shows the F-concentration dependence of the obtained $\nu _c\sqrt{1 +  \frac{\eta ^2}{9}}$ at three representative temperatures. For all cases, 
 $\nu _c\sqrt{1 +  \frac{\eta ^2}{9}}$ changes progressively as F-content increases but shows a crossover at a certain F-concentration $x_s$ that is temperature dependent.
%
%
This result together with the fact  that the lattice constant obtained by XRD also changes continuously as $x$ increases \cite{Supple} assure that the doping does increase with increasing $x$.
Figure \ref{nu_all}(b) shows the temperature dependence of $\nu _c\sqrt{1 +  \frac{\eta ^2}{9}}$ for $x$ = 0.3, 0.5 and 0.55. For the $x$ = 0.3 and 0.5 compounds, this quantity is temperature independent. By contrast, for  $x$ = 0.55, 
there is a clear transition taking place at a temperature between 250 and 200 K.  The middle point of the transitions is defined as $T_s$. The $x_s$ and $T_s$ are plotted in Fig. \ref{phase_diagram}(a).
For $x\geq$0.6, the above quantity  is $T$-independent again below room temperature~\cite{Supple}.
Since it has been established that $\nu_c$ increases linearly with increasing doping \cite{Oka} and decreases smoothly with decreasing temperature~\cite{Supple}, the present  results indicate that $\eta$ undergoes an abrupt increase when crossing  $x_s$ or $T_s$. It is worthwhile pointing out that the $T_s$ line extrapolates to absolute zero at  $x\sim$0.5, where
the temperature dependence of the electrical resistivity shows a critical behavior of $\rho - {{\rho }_{0}}\propto {{T}}$ . 

The abrupt change of $\eta$ crossing  $x_s$ is further corroborated  by the measurements at the La site.  Figure \ref{nu_all}(c) shows the  $^{139}$La ($I$ = 7/2) NQR transition lines for $x$ = 0.3 and 0.55. As seen in the inset to Fig. \ref{nu_all}(c),  the three transition lines are equally separated when $\eta$ = 0. For finite $\eta$, however, the three lines are not equally separated; in particular, the lowest ($m$ = $\pm$1/2 $\leftrightarrow$ $\pm$3/2) transition  line largely shifts to a higher frequency as $\eta$ increases. Such difference is exactly found between the $x$ = 0.3 and $x$ = 0.55 compounds. Our results reveal    that $\eta$ = 0 for $x$ = 0.3 but $\eta$ = 0.21 for  $x$ = 0.55. 
Thus, above the new dome, there is a  structural phase transition from a four fold symmetry at high temperature to a lower symmetry at low temperature.


\begin{figure}[htbp]
\includegraphics[width=7cm]{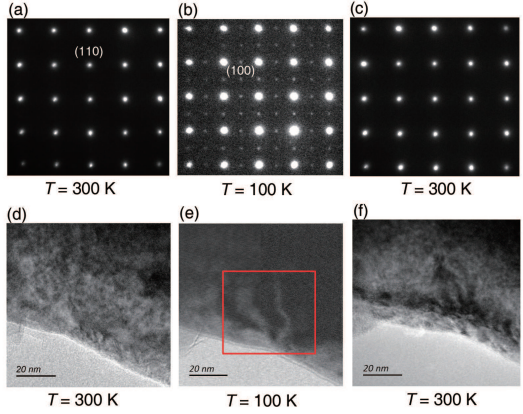}
\caption{Electron diffraction patterns of LaFeAsO$_{0.45}$F$_{0.55}$ during a heat cycling of 300 K (a) $\rightarrow$ 100 K (b) $\rightarrow$ 300 K (c). 
Additional Bragg spots appear at the systematic (100) positions, indicating a   change in crystal symmetry.
(d-f) show the microstructures during the cycling.  Domain structure appeared at $T$=100 K as seen in the area marked by the red rectangle, but disappeared after heating back to 300 K.}
\label{heatcycling}
\end{figure}

\begin{figure}[htbp]
\includegraphics[width=8.5cm]{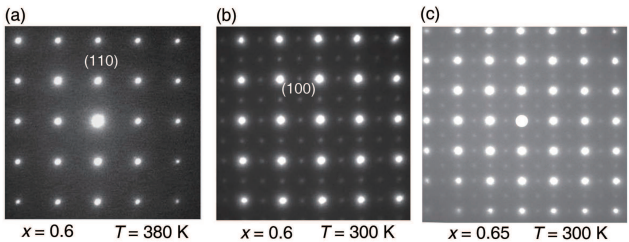}
\caption{Electron diffraction patterns of LaFeAsO$_{0.4}$F$_{0.6}$ and LaFeAsO$_{0.35}$F$_{0.65}$. (a-b) TEM data for  $x = 0.6$ at $T$ = 380 K and 300 K. The images were taken along the [001] zone-axis direction. 
(c) The TEM image for $x = 0.65$ at $T$ = 300 K. The crystal symmetry of this compound is also  of orthorhombic already at room temperature.}
\label{TEM5}
\end{figure}

Furthermore, we have directly confirmed the structural transition by Transmission Electron Microscope (TEM). Figure \ref{heatcycling}(a) shows the [001] zone-axis electron diffraction patterns for  LaFeAsO$_{0.45}$F$_{0.55}$ taken at $T$ = 300 K, which indicates that the crystal structure is tetragonal with the space group of $P$4/$nmm$. At $T$ = 100 K, however, additional spots appear at  (100) positions, as seen in Fig. \ref{heatcycling}(b). These features indicate unambiguously that the symmetry is lowered at low temperature.
Furthermore, domain structures appear at low temperature as shown in Fig. \ref{heatcycling}(e) which are absent at $T$ = 300 K. These structures are due to twinning domains energetically preferred by structural relaxation following the phase transition.

Figure \ref{heatcycling}(a-c) shows the TEM results for $x$ = 0.55 during a heat cycling of 300 K $\rightarrow$ 100 K $\rightarrow$ 300 K, and (d-f) the corresponding microstructures. After cooling to 100 K and measuring the TEM images there, the sample was heated back to 300 K and measured again. We find that  the tetragonal symmetry is recovered and the TEM data are reproducible. These results indicate that the additional spots  and the domain structure found at $T$ = 100 K  are  intrinsic properties due to the structural transition. 
Around the structural phase transition temperature, we have taken TEM pictures in a step of 5 K and  found no hysteresis during the heat cycling.
It is noted that the observed changes both in the Bragg spots and in the microstructure   agree completely with the cases of parent compounds LaFeAsO and NdFeAsO where a tetragonal-to-orthorhombic  structural  transition takes place at $T_{\rm s}$ = 135 K \cite{YangHX}.
Also, we have confirmed that the compounds with $x$=0 and 0.1 synthesized under the same conditions as those for $x\geq$0.3 do have  C4 symmetry at room temperature
, so the C4 symmetry breaking below $T_{\rm s}$ for $x>$0.5 is not due to the high temperature and  high pressure treatment.

Evidence for the structural  transition  was also obtained for $x$ = 0.6 as shown in Fig. \ref{TEM5}(a-b).
At $T$ = 300 K, the (100) spots and domain structures indicating  C4-symmetry breaking was found.
 Upon heating from room temperature,  the (100) spots  disappeared at $T$ = 380 K. We therefore determined  $T_s$ = 380$\pm$5 K for this composition and plotted this point in Fig. \ref{phase_diagram}(a).
Figure \ref{TEM5}(c) shows the TEM taken at $T$ = 300 K for $x$ = 0.65. As for the $x$ = 0.6 compound discussed above, the C4 symmetry is also broken already at $T$ = 300 K for this composition. The TEM results are in agreement with the NMR data of $T_s>$  300 K for $x$ $\geq$ 0.6 .


\begin{figure}
\includegraphics[width=9cm]{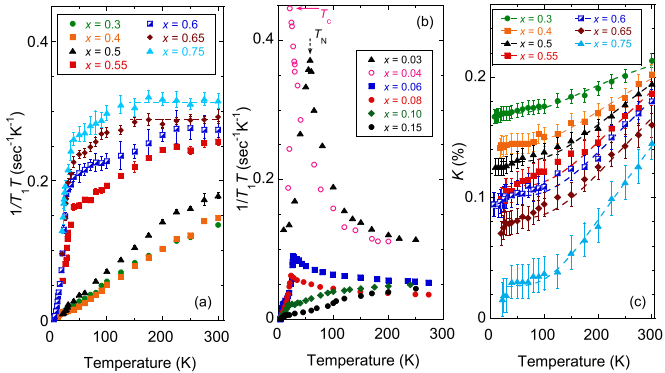}
\caption{ The quantity 
$1/T_{1}T$ and the Knight shift.
(a) The variation of $1/T_{1}T$  for $x\geq$0.3. The broken lines 
show a $1/T_{1}T$ = const. relation.   (b) Data for $x \leq$0.15 taken from Ref. \cite{Oka}. 
(c) The temperature dependence of  the Knight shift $K$  of $H_{0}$ $\|$ $ab$ for $x\geq$0.3. 
}
\label{T1TK3}
\end{figure}

The tetragonal-to-orthorhombic structural transition and the associated electronic nematicity (in-plane anisotropy) has been a main focus in the study of the mechanism for high-$T_c$ superconductivity in iron-pnictides.
In LaFeAsO  or underdoped BaFe$_{2-x}$M$_x$As$_2$ (M = Ni, Co), a structural transition   takes place above the magnetic order temperature $T_N$ \cite{Dai,ZhouRui,Safet}~. Below $T_s$, many physical properties exhibit nematicic behavior  \cite{Chu,Imai-LaFeAsO}~.
It has been proposed  that the structural transition, the electronic nematicity, the magnetic transition and the  superconductivity are inter-related \cite{Fernandes-NatPhys}~. 

Currently,   there are two schools of theories. One is that the structural transition and the resulting electronic nematicity are
directly driven by strong low-energy magnetic  fluctuations \cite{Fernandes-2}.
However, 
our system is far away from a magnetic ordered phase, and there is no indication of low-energy magnetic fluctuations. 
Figure \ref{T1TK3} (a) shows the temperature dependence of the quantity 1/$T_1T$, with comparison to that for $x\leq$ 0.15 shown in Figure \ref{T1TK3} (b) \cite{Oka}.
 Figure \ref{T1TK3} (c) shows the Knight shift $K$ for the compounds of the new dome. 
 For a conventional metal,  both 1/$T_1T$ and $K$ are temperature-independent.
 The first intriguing aspect seen from Fig. \ref{T1TK3} (a) is that  $1/T_1T$ appears to fall into two groups, one for $x\leq$ 0.5 and the other for $x>0.5$, which is consistent with the structural phase boundary  shown in Fig. \ref{phase_diagram}.
Secondly,  the two quantities  1/$T_1T$ and $K$ for $x\geq$ 0.3 show a similar trend in the temperature variation; they decrease with decreasing temperature, suggesting that both of them  are dominated by  the density of states at the Fermi level. 
This is  in sharp contrast to the low-doped compounds ($x\leq$0.1) where  1/$T_1T$ increases rapidly upon cooling \cite{Oka,Nakai}, due to the development of low-energy spin fluctuations. 
As can be seen in  \ref{T1TK3} (b), when there is a Neel order, as is the case for  $x$ = 0.03, 1/$T_1T$ increases with decreasing temperature, forming a sharp peak at the Neel temperature $T_{\rm N}$, due to a critical slowing down of the magnetic moments.
When the magnetic order is suppressed by doping but magnetic (spin) fluctuations are present,   1/$T_1T$ follows a Curie-Weiss type of variation before superconductivity sets in \cite{Oka}, which is the case for $x$ = 0.04, 0.06 and 0.08. However, non of these features are observed in the compounds of $x \geq$ 0.3.

\begin{figure}
\includegraphics[width=4cm]{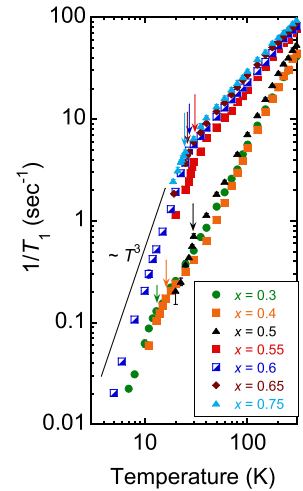}
\caption{ The  temperature dependence of the $^{75}$As spin-lattice relaxation rate $1/T_1$.  The arrows indicate $T_c$ under the magnetic field of $H\sim$12 T measured by the $in situ$ NMR coil. 
The straight line is a guide to the eyes to indicate the  $1/T_1\propto T^3$ relation.}  
 \label{1_T1_all}
 \end{figure}

The second school of theories is that
 orbital order or fluctuations  are responsible for the nematic structural transition and the superconductivity 
\cite{Kontani,Chen,Lee,Philip,Yamase}. However, the proposed orbital fluctuations with finite momentum can be enhanced only when low-energy spin fluctuations exist \cite{Kontani,Lee}.

Thus, in the absence of low-energy spin fluctuations, the newly-discovered second dome with $T_c$ even higher than the first dome, and  the structural transition above $T_c$,  pose a challenging   issue.
In this regard, the $T$-linear electrical resistivity seen around $x_{opt}$ may provide a clue. Quantum fluctuations of an order parameter can lead to non-Fermi liquid behavior of the electrical resistivity \cite{Hertz}. In particular, a $T$-linear resistivity   was often taken as a fingerprint of two-dimensional quantum magnetic fluctuations \cite{Moriya_SCR}~. In the present case where  low-energy magnetic fluctuations are absent, it is tempted to consider that the observed $T$-linear resistivity is  caused by the fluctuations of widely-searched electronic nematic order such as orbital order. 
In fact, a similar $T$-linear resistivity was found in BaFe$_{2-x}$Ni$_{x}$As$_{2}$ ($x$=0.14) which is far away from the magnetic quantum critical point ($x$=0.10) and  $T_s$ extrapolates to zero there \cite{ZhouRui}.
 To our knowledge, the temperature dependence of the electrical resistivity due to  orbital fluctuations has not been theoretically worked out  yet. Our work indicates that exploring the origin of the $T$-linear electrical resistivity is an important task 
since the issue  may further be related  to the driving force of the structural transition as well as  to the mechanism of superconductivity in the whole iron-pnictides family. Also, future study of the  nematicity  in the physical properties 
for this system will greatly help move the research field further ahead.

Finally, we briefly touch on the properties of the superconducting state. Figure \ref{1_T1_all} shows the temperature dependence of the spin-lattice relaxation rate $1/T_1$ for the new dome. For all F-concentrations, there is  a clear and sharp decrease below $T_c$, which assures the bulk nature of the superconductivity. The detailed temperature variation below $T_c$ seems to be similar between $x$ = 0.6 in the new dome and $x$ = 0.08 in the first dome \cite{Supple}.

Before closing, we emphasize that the phase diagram  discovered here resembles that of heavy fermion compound CePd$_2$Si$_2$ where a $T$-linear resistivity was also found at the top of the dome \cite{Lonzarich}~. However, the $T_N$ line in CePd$_2$Si$_2$ is replaced by $T_s$ line in our case.
Therefore,  the new superconductivity dome  reported here may represent a new paradigm of superconductivity.
 Our finding may also shed light on the physics of heavy fermion materials such as CeCu$_2$Si$_2$ where two superconducting phases, one close to magnetism and the other far away from magnetism,  were reported \cite{Yuan}~.
In the context of iron pnictides, the heavily F-doped LaFeAsO$_{1-x}$F$_x$ system made by the high-pressure synthesis technique also represents a unprecedented class of superconductors.  Previously, hydrogen substitution for oxygen up to 53\% was reported to dope electron efficiently and induce two superconducting regions that merge into one at high pressure \cite{Iimura}~. However,  all the  superconducting regions there have a tetragonal crystal structure with C4 symmetry and the  superconducting phase at the high-doping region is in close proximity to an adjacent magnetic ordered phase at $x\geq$0.53 \cite{Kojima}. 

In conclusion,  we have reported the discovery of a new superconductivity dome without low-energy magnetic fluctuations in LaFeAsO$_{1-x}$F$_{x}$ with 0.25$\leq x \leq$0.75, where the maximal critical temperature $T_c$ at $x_{opt}$ = 0.5$\sim$0.55 is even higher than that at $x \leq$ 0.2. By nuclear magnetic resonance and Transmission Electron Microscopy, we demonstrated that a C4 rotation symmetry-breaking structural transition takes place for $x>$ 0.5 above $T_c$.  Our results point to a new paradigm of high temperature superconductivity, and    suggest that 
there may be a new route to high temperature superconductivity and that superconductors with further higher $T_c$  await discovery.

\begin{acknowledgments}
We thank   Z. Li   for assistance in some of the measurements,  T. Xiang, L. L. Sun, S. Onari, H. Ikeda and H. Kontani for helpful discussion. This work was partially supported by CAS's Strategic Priority Research Program, No. XDB07020200,  National Basic Research Program of China, Nos. 2012CB821402,   2011CBA00109 and 2011CBA00101, and by NSFC Grant No 11204362.
\end{acknowledgments}



\begin{thebibliography}{99}
\bibitem{Supple}
More data will be presented in a separate paper.
\end{thebibliography}

%
%
%


\end{document}